\documentclass[pra,twocolumn,showpacs,amsmath]{revtex4}

\usepackage{graphicx}
\begin{document}

\title{Validity of adiabaticity in Cavity QED}
\author{J. Larson and S. Stenholm \\
Physics Department\\
Royal Institute of Technology (KTH)\\
Albanova, Roslagstullsbacken 21\\
SE-10691 Stockholm, Sweden}

\begin{abstract}
This paper deals with the concept of adiabaticity for fully quantum mechanically cavity QED models. The physically interesting cases of Gaussian and standing wave shapes of the cavity mode are considered. An analytical approximate measure for adiabaticity is given and compared with numerical wave packet simulations. Good agreement is obtained where the approximations are expected to be valid. Usually for cavity QED systems, the large atom-field detuning case is considered as the adiabatic limit. We, however, show that adiabaticity is also valid, for the Gaussian mode shape, in the opposite limit. Effective semiclassical time dependent models, which do not take into account the shape of the wave packet, are derived. Corrections to such an effective theory, which are purely quantum mechanical, are discussed. It is shown that many of the results presented can be applied to time dependent two-level systems.     
\end{abstract}


\maketitle

\section{Introduction}

The concept of adiabaticity dates back to the early days of quantum mechanics and has, in spite of its simplicity, drawn attention ever since. In this paper we investigate adiabaticty within cavity quantum electrodynamics (QED) models, namely for extended Jaynes-Cummings models \cite{jc}. A two-level atom interacts with a quantized cavity mode, where both atomic center-of-mass momentum and position are taken into account. The adiabatic theorem states that, in the limit of an infinitely slowly changing Hamiltonian $H(t)$ time $t$ may be seen as a parameter and $H(t)$ can be instantaneously diagonalizable and no population transfer will take place between the instantaneous eigenstates \cite{adiabat}. If the change governs a closed loop in 'parameter' space, it has been shown that the eigenstates do not only accumulate the standard dynamical phase, but also a geometric one \cite{berry}. Adiabatic methods have also been suggested for realizing quantum information processing, causing the parameters to be stable against fluctuations \cite{adcomp}. Recently, an inconsistency of the criteria for the theorem  has been proposed \cite{adcond}, which has led to new insight into the theorem and several commenting papers \cite{adreplay}. The adiabatic theorem is usually associated with an explicitly time dependent problem. However, the phenomena has a broader spectrum of application, for example: adiabatic elimination of internal states \cite{adel1,adel2} and the Born-Oppenheimer approximation \cite{born}, and within wave packet propagation. Adiabatic elimination gives effective models when the variations of some slowly varying quantities are neglected from the equations of motion. For wave packet propagation, which will be considered in this paper, the rate of change is governed by the motion of the wave packet. This is easily seen for particle with mass $m$ and kinetic energy much larger than the interaction energy; it is often legitimate to replace momentum and position in the Hamiltonian by their classical counterparts, $p\rightarrow mv$ and $x\rightarrow vt$. In this way an effective time dependent model is obtained \cite{jonas1,jonas2}, which does, however, not take into account the shape of the wave packet. It is known that, for small kinetic energies, treating $x$ and $p$ as classical variables may not be justified, and a full quantum analysis must be carried out \cite{slow,jonas6}. 

We introduce a dynamic and adiabatic frame in sections \ref{sec2} and \ref{sec3} and give a criterion for adiabatic evolution of the wave packets in section \ref{sec4}. This measure is approximated to first order and checked with numerical wave packet propagation. Good agreement is obtained, when expected according to the approximations. Higher order corrections, for example when the shape of the wave packet is taken into account, are discussed in section \ref{sec7}. The generally associated adiabatic limit $|\Delta|\gg G(x)$, where $G(x)=\sqrt{n}g(x)$ is the atom-field coupling constant multiplied by the square root of the photon number,  reproduces adiabaticity in our model. It is, however, found that the opposite limit, $|\Delta|\ll G(x)$, can also give an adiabatic process. This is counterintuitive from what is formally believed, and is therefor discussed in detail in the paper. Two special cases of a Gaussian and a standing wave mode shape are considered separately in subsections \ref{ssec51} and \ref{ssec52} for a theoretical analysis and in \ref{ssec61} and \ref{ssec62} for numerical simulations. In the former Gaussian model, with $|\Delta|\ll G(x)$, an adiabatic evolution is obtained for all $x$, while for the standing wave mode with wave length $\lambda$, the adiabaticity breaks down around the points $x=\pi n/q$, where $n=...,-2,-1,0,1,2,...$ and $q=2\pi/\lambda$. As the approximated adiabaticity criteria have identical form as the adiabaticity criteria obtained in time dependent models, much of the analysis in this paper also applies to such systems. This is discussed in section \ref{sec8}, where we also give a systematic way to derive a semiclassical time dependent Hamiltonian from time independent fully quantum mechanical ones.  

\section{The dynamic frame}\label{sec2}

We can easily write down the Hamiltonian describing a single two-level atom
passing an electromagnetic field confined to a cavity. In addition to the
customary Jaynes-Cummings terms, we have to add the energy of the free atom
and the spatial variation it feels from the caity. With $\hbar =1$, we obtain%
\begin{equation}
H_{1}=\frac{p^{2}}{2m}+\omega a^{\dagger }a+\frac{\Omega }{2}\sigma
_{3}+H_{2}(x),  \label{a1}
\end{equation}%
where the cavity mode interaction is%
\begin{equation}
H_{2}(x)=g(x)\left( a\sigma ^{+}+a^{\dagger }\sigma ^{-}\right) .
\label{a1a}
\end{equation}%
Here $\Omega $ is the energy separation of the two-level system, $\omega $
is the energy of the photon created by $a^{\dagger },$ and the $\sigma :s$
are the Pauli matrices. The excitation number operator%
\begin{equation}
N=a^{\dagger }a+\frac{1}{2}\sigma _{3}  \label{a2}
\end{equation}%
commutes with the Hamiltonian and thus we find separation of the blocks of
the state space labelled by the excitation number%
\begin{equation}
 \tilde{\Phi}^{n} (p)=\tilde{\Phi}_{+}(p)\mid +,n-1\rangle +\tilde{\Phi}_{-}(p)\mid
-,n\rangle .  \label{a3}
\end{equation}%
The translational degree of freedom is found in the coefficients $\tilde{\Phi}_{\pm }.$

We may, however, separate an arbitrary multiple $\Lambda $ of $N$ because of
its commutation with the Hamiltonian. We define the states%
\begin{equation}
 \Phi ^{n} =\exp \left( -i\Lambda Nt\right)  \tilde{\Phi}%
^{n}  \label{a4}
\end{equation}%
evolving with the Hamiltonian%
\begin{equation}
\begin{array}{l}
H=H_{1}-\Lambda N \\ 
\\ 
=\frac{p^{2}}{2m}+\left( \omega -\Lambda \right) a^{\dagger }a+ 
\frac{\Omega-\Lambda }{2}\sigma _{3}+V(x).%
\end{array}
\label{a5}
\end{equation}%
These are the generalisations of going to a rotating frame. We now have two
important cases:\\

\textit{Case} 1: $\Lambda =\omega .$

\textit{Case} 2: $\Lambda =\Omega .$\\ \\
If we write the equations for the state in the general form%
\begin{equation}
i\frac{\partial }{\partial t}\left[ 
\begin{array}{c}
\tilde{\Phi}_{+} \\ 
\tilde{\Phi}_{-}%
\end{array}%
\right] =\frac{p^{2}}{2m}\left[ 
\begin{array}{c}
\tilde{\Phi}_{+} \\ 
\tilde{\Phi}_{-}%
\end{array}%
\right] +\left[ 
\begin{array}{cc}
\varepsilon _{+} & G \\ 
G & \varepsilon _{-}%
\end{array}%
\right] \left[ 
\begin{array}{c}
\tilde{\Phi}_{+} \\ 
\tilde{\Phi}_{-}%
\end{array}%
\right]  \label{a6}
\end{equation}%
we find with 
\begin{equation}
\begin{array}{ccc}
\Delta =\Omega -\omega & ; & G=g(x)%
\sqrt{n}%
\end{array}
\label{a7}
\end{equation}%
the results%
\begin{equation}
\begin{array}{c}
{\textit Case\,\, 1:}\\ 
\varepsilon _{\pm }=\pm \frac{\Delta }{2}%
\end{array}
\label{a8}
\end{equation}%
and%
\begin{equation}
\begin{array}{c}
{\textit Case\,\, 2:} \\ 
\varepsilon _{+}=-\Delta \left( n-1\right) \\ 
\varepsilon _{-}=-\Delta n.%
\end{array}
\label{a9}
\end{equation}

\section{The adiabatic frame}\label{sec3}

If we now define the two-level potential from (\ref{a6}) as%
\begin{equation}
V(x)=\left[ 
\begin{array}{cc}
\varepsilon _{+} & G \\ 
G & \varepsilon _{-}%
\end{array}%
\right]  \label{a10}
\end{equation}%
we may introduce the unitary transformation 
\begin{equation}
U=\left[ 
\begin{array}{cc}
\cos \theta & \sin \theta \\ 
-\sin \theta & \cos \theta%
\end{array}%
\right] ,  \label{a11}
\end{equation}%
such that the adiabatic potential%
\begin{equation}
UVU^{\dagger }=W\equiv \left[ 
\begin{array}{cc}
\Delta _{+} & 0 \\ 
0 & \Delta _{-}%
\end{array}%
\right]  \label{a12}
\end{equation}%
is diagonal. However, since
\begin{equation}
U(x)pU^{\dagger }(x)=p-\sigma _{2}\partial \theta ,  \label{a20}
\end{equation}%
where%
\begin{equation}
\partial \theta \equiv \frac{d\theta (x)}{dx},  \label{a21}
\end{equation}%
the transformed Hamiltonian is non-diagonal. Explicitly we have
\begin{equation}\label{Htrans}
\tilde{H}=H_{ad}+H_{cor},
\end{equation}
with the adiabatic part
\begin{equation}\label{adH}
H_{ad}=\frac{p^2}{2m}+\left[\begin{array}{cc}\Delta_+ & 0 \\ 0 & \Delta_-\end{array}\right],
\end{equation}
and the non-adiabatic correction part
\begin{equation}\label{nonadH}
H_{cor}=\frac{1}{2m}\left[\begin{array}{cc}(\partial\theta)^2 & -2(\partial\theta) p-\partial^2\theta \\ 2(\partial\theta) p+\partial^2\theta & (\partial\theta)^2\end{array}\right].
\end{equation}
The transformed Hamiltonian is given in the adiabatic basis
\begin{equation}\label{adbasis}
\begin{array}{l}
|\uparrow\rangle_n=\cos\theta|n,+\rangle+\sin\theta|n+1,-\rangle, \\ \\
|\downarrow\rangle_n=-\sin\theta|n,+\rangle+\cos\theta|n+1,-\rangle,
\end{array}
\end{equation}
and for a given $n$, a general state is written, in the adiabatic basis, as
\begin{equation}\label{adwavepacket}
\Psi(x,t)=\Psi_\uparrow(x,t)|\uparrow\rangle_n+\Psi_\downarrow(x,t)|\downarrow\rangle_n.
\end{equation}
The eigenvalues are%
\begin{equation}
\Delta _{\pm }=\frac{\varepsilon _{+}+\varepsilon _{-}}{2}\pm \sqrt{\left( 
\frac{\Delta \varepsilon }{2}\right) ^{2}+G^{2}},  \label{a15}
\end{equation}%
where%
\begin{equation}
\Delta \varepsilon =\varepsilon _{+}-\varepsilon _{-},  \label{a14}
\end{equation}%
and the diagonalisation gives the result%
\begin{equation}
\tan 2\theta =\frac{2G}{\Delta \varepsilon },  \label{a13}
\end{equation}%

The interesting observation is that in both \textit{Case} 1 and \textit{Case}
2, we find%
\begin{equation}
\Delta \varepsilon =\Delta ,  \label{a16}
\end{equation}%
which shows that the energy splitting (\ref{a15}) in the adiabatic frame
becomes the same. On the other hand, the splitting is centered on different
positions because%
\begin{equation}
\begin{array}{lll}
\bar{\varepsilon}\equiv \frac{\varepsilon _{+}+\varepsilon _{-}}{2} & =0; & 
({\textit Case}\,\, 1) \\ 
&  &  \\ 
& =-\Delta \left( n-\frac{1}{2}\right) ; & ({\textit Case}\,\, 2).%
\end{array}
\label{a17}
\end{equation}

For large detuning $\Delta ,$ we find from (\ref{a15})%
\begin{equation}
\Delta _{\pm }=\bar{\varepsilon}\pm \frac{\Delta }{2}\pm \frac{G^{2}}{\Delta 
}.  \label{a18}
\end{equation}%
As expected, the levels are pushed apart for both signs of the detuning $%
\Delta$. In this limit, the atomic motion is taking place on the effective
potential surface%
\begin{equation}
V_{eff}=\frac{g(x)^{2}}{\Delta }n.  \label{a19}
\end{equation}%
We see that the spatial dependence of the mode structure is enhanced by the
number of photons in the cavity mode. This limit is the one usually intended
in the discussion of adiabatic following through the cavity-induced potential.

When the detuning goes to zero, $\Delta \rightarrow0,$ the adiabatic potentials follow
the shape of the mode functions, (\ref{a15}) becomes%
\begin{equation}
\Delta _{\pm }=\frac{\varepsilon _{+}+\varepsilon _{-}}{2}\pm g(x)\sqrt{n}.
\label{a19a}
\end{equation}

However, formally the adiabatic diagonalisation can be performed for all
values of $\Delta $, and we want to enquire whether the result (\ref{a15})
has any validity beyond the customary large detuning limit?

\section{Validity of adiabaticity}\label{sec4}

The correction introduced by the adiabatic approximation follows from the
transformation of the kinetic energy due to the position-dependent
diagonalizing operator, seen in eqs. (\ref{a20}) and (\ref{a21}). We observe that if the angle $\theta $ were equipped with independent
dynamics, this would be analogous to a gauge field situation. The adiabatic approximation holds when $H_{ad}$ dominates the evolution compared to $H_{cor}$, thus when the distance between the adiabatic eigenvalues is much larger than the off-diagonal term. This gives a condition for adiabaticity accordingly
\begin{equation}
\left\vert\langle\Delta_+\rangle_i-\langle\Delta_-\rangle_i\right\vert\gg\frac{1}{2m}\left\vert\langle 2(\partial\theta)p\rangle_i+\langle\partial^2\theta\rangle_i\right\vert.  \label{a24}
\end{equation}%
Here we have averaged over the adiabatic wave packet $\Psi_i^{ad}(x,t)$ defined as the initial wave packet propagated by the adiabatic Hamiltonian (\ref{adH})
\begin{equation}\label{adstate}
\Psi_i^{ad}(x,t)\!=\!\exp\!\!\left[\!-i\left(\frac{p^2}{2m}+\!\Delta_\pm\!\right)\!t\!\right]\!\Psi_i(x,0),\hspace{0.4cm}i=\downarrow,\,\uparrow,
\end{equation}
where it is understood that $\Delta_+$ corresponds to the $i=\uparrow$ and vice versa for $\Delta_-$. Hence for any operator $h(x,p)$:
\begin{equation}
\langle h(x,p)\rangle_i\!=\!\!\int\! dx\Psi_i^{ad*}(x,t)h(x,p)\Psi_i^{ad}(x,t),\hspace{0.4cm}i=\downarrow,\,\uparrow\!.
\end{equation}
For a more general initial state, any linear combination of $|\downarrow\rangle$ and $|\uparrow\rangle$ states ($n=1$), we define the adiabaticity parameter as
\begin{equation}\label{adpara}
\mathcal{A}_t\!\!=\!\!\frac{1}{2m}\!\!\left(\!\!N_\downarrow^0\!\frac{\left\vert\!\langle2(\partial\theta)p\rangle_\downarrow\!+\!\langle\partial^2\theta\rangle_\downarrow\!\right\vert}{\left\vert \langle\Delta_+\rangle\downarrow-\langle\Delta_-\rangle\downarrow\right\vert}\!\!+\!\!N_\uparrow^0\!\frac{\left\vert\!\langle2(\partial\theta)p\rangle_\uparrow\!+\!\langle\partial^2\theta\rangle_\uparrow\!\right\vert}{\left\vert \langle\Delta_+\rangle\uparrow-\langle\Delta_-\rangle\uparrow\right\vert}\!\!\right)\!,
\end{equation}
where $N_i^0$ is the initial population in the state $|i\rangle$, $i=\downarrow,\,\uparrow$. Thus $\mathcal{A}_t\ll1$ implies an adiabatic evolution for the particular initial state. In time dependent systems, the adiabaticity constrain is obtained directly by instantaneous diagonalization of the Hamiltonian. In our model, however, this is not the case; the diagonalization gives $\Psi^{ad}(x,0)_i$, but its time evolution is obtained from (\ref{adstate}). 

To obtain some insight into the adiabaticity parameter, some approximations are in order. It is reasonable to assume that the main non-adiabatic contributions, due to the evaluation with the Hamiltonian $H_{cor}$ (\ref{nonadH}), comes from the off-diagonal term $\partial\theta p$. We will therefor neglect the term $\partial^2\theta$ at the moment and discuss its influence in section \ref{sec7}. For a particle moving with a high velocity, we make the assumption that $p\approx p_0$ for the two wave packets $\Psi_{\uparrow,\downarrow}^{ad}(x)$, where $p_0$ is the initial velocity. The position is thus given by $x=p_0t/m+x_0$ ($x_0$ is the initial position and is taken to be the same for both states). We return to this approximation in section \ref{sec7}. To evalute (\ref{a21}), we take the derivative of (\ref{a13}) to obtain%
\begin{equation}
\frac{2\partial \theta }{\cos ^{2}2\theta }=\left( 1+\tan ^{2}2\theta
\right) 2\partial \theta =\frac{2\partial G}{\Delta }.  \label{a22}
\end{equation}%
This gives%
\begin{equation}
\partial \theta =\frac{\Delta \partial G}{\Delta ^{2}+4G^{2}}=\frac{\Delta \,%
\sqrt{n}\,\partial g}{\Delta ^{2}+4ng^{2}}.  \label{a23}
\end{equation}%
Summarizing the above and using eq. (\ref{adpara}), we find the approximate expression
\begin{equation}\label{adparaapp}
\mathcal{A}^0(x)=\left\vert\left(\frac{p_0}{m}\right)\frac{\Delta\sqrt{n}\partial g}{\left(\Delta^2+4ng^2\right)^{3/2}}\right\vert.
\end{equation}
As it stands, $\mathcal{A}^0(x)$ is given as a function of the parameter $x$, which is, however, equivalent to expressing it as a function of time $t$ by the substitution $x=p_0t/m+x_0$. The expression (\ref{adparaapp}) has got two limiting behaviours:
For $\mid \Delta \mid \gg 2g\sqrt{n}$ we obtain%
\begin{equation}
\mathcal{A}^0(x)\approx\left\vert\left(\frac{p_0}{m}\right)\frac{\sqrt{n}\partial g}{\Delta^2}\right\vert.  \label{a25}
\end{equation}%
This gives the condition for the validity of the ordinary adiabatic result (\ref%
{a19})$.$ In the opposite limit $\mid \Delta \mid \ll 2g\sqrt{n},$ we find%
\begin{equation}
\mathcal{A}^0(x)\approx\left\vert\left(\frac{p_0}{m}\right)\frac{\Delta\sqrt{n}\partial g}{8ng^3}\right\vert.   \label{a26}
\end{equation}%
This result is remarkable because it favours a high level of cavity
excitation and it disappears for vanishing detuning. This seems to indicate
that the adiabatic approximation may also be valid for resonance between the
atoms and the cavity mode. However, in order to see how this comes about, we need to
investigate the behaviour in some special cases. 

\section{Special cases}\label{sec5}

\subsection{Gaussian mode shape}\label{ssec51}

If the atoms traverse the mode structure in a
transverse direction of a Fabry-Perot cavity, the amplitude seen is of Gaussian type. Thus we set%
\begin{equation}
g(x)=\frac{A}{\sqrt{2\pi }a}\exp \left( -\frac{x^{2}}{2a^{2}}\right) .
\label{a27}
\end{equation}%
For this coupling, the adiabaticity criterion $\mathcal{A}^0(x)\ll1$ reads
\begin{equation}
\left\vert \left(\frac{p_0}{m}\right)\frac{\Delta \sqrt{n}}{\left(\Delta
^{2}+4ng^{2}(x)\right)^{3/2}}\left( \frac{x}{a}\right) g(x)\right\vert \ll1.  \label{a28}
\end{equation}%
For large detunings, because $x\lesssim a,$ this becomes approximately%
\begin{equation}
\left\vert\left(\frac{p_0}{m}\right)\frac{Ap_0\sqrt{n}}{a \Delta
^2 }\right\vert\ll1.  \label{a29}
\end{equation}%
Like (\ref{a25}) this is the ordinary condition for adiabatic motion; in
particular, it is less valid for high level of excitation.

In the opposite limit, $\Delta \rightarrow 0$, we find%
\begin{equation}
1\gg \left\vert \left( \frac{x}{a}\right) \left(\frac{p_0}{m}\right)\frac{%
\Delta }{8ng^{{2}}(x)}\right\vert \sim \left\vert\frac{p_0}{m}\frac{ \Delta
}{an}x\right\vert \exp \left( \frac{x^{2}}{a^{2}}\right) ,
\label{a30}
\end{equation}%
This seems to vanish for $\Delta =0$ but, on the other hand, it blows up for 
$x>a$. 

\begin{figure}[ht]
\centerline{\includegraphics[width=8cm]{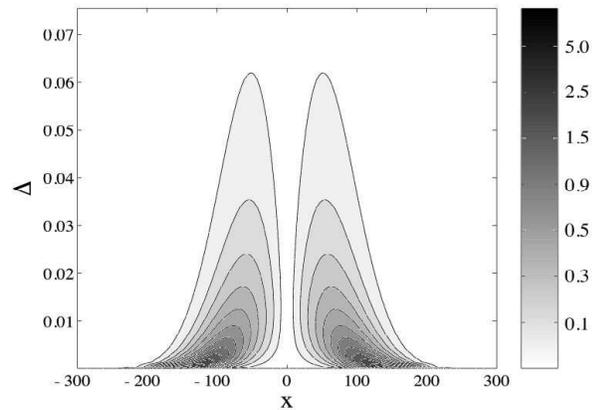}}
\caption[adiabaticpara1]{The adiabaticity parameter $\mathcal{A}^0(x)$ as function of $x$ and $\Delta$ in scaled dimensionless units (starting from $\Delta=0.0001$) for a Gaussian coupling (\ref{a27}) with amplitude $A=1$, the pulse width $a=50$ and the initial momentum $p_0=10$. \label{fig1} }
\end{figure}

In deriving (\ref{a30}) we assumed $|\Delta|\ll\sqrt{n}g(x)$, which will clearly break down for values on $x$ large enough, say $x>a$. This is exactly when the adiabaticity parameter $\mathcal{A}^0(x)$ seems to blow up according to (\ref{a30}). For such values this approximation is not likely to be valid, and we need to explore the bahaviour of $\mathcal{A}^0(x)$ for large $|x|$ and small $\Delta$ in more detail. Is $\mathcal{A}^0(x)$ approaching 0 smoothly or discontinuously when $\Delta\rightarrow0$? In fig. \ref{fig1} we plot $\mathcal{A}^0(x)$ as function of $x$ and $\Delta$ (starting from $\Delta=0.0001$). In the following we consider {\it Case} 1: $\Lambda=\omega$, and use scaled parameters such that $\hbar=m=1$ and put the photon number $n=1$ in all of the numerical calculations. 

\begin{figure}[ht]
\centerline{\includegraphics[width=8cm]{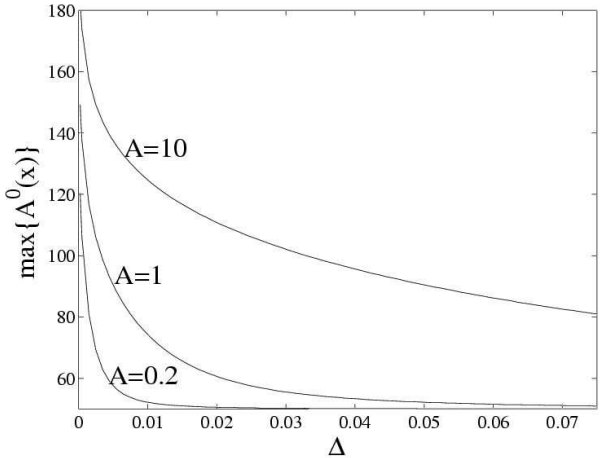}}
\caption[maxA]{The location of the maximum $\mathrm{max}\{\mathcal{A}^0(x)\}$ as function of the dimensionless parameter $\Delta$ for three different couplings. The pulse width is again $a=50$ and the initial momentum $p_0=10$. \label{fig2} }
\end{figure}

We note that $\mathcal{A}^0(x)$ has two maxima for a given detuning $\Delta$, symmetrically situated around $x=0$. The value of this maximum is clearly increasing for small detunings, but it is also moving towards $|x|\gg a$. Thus, in the very limit of vanishing detuning, the adiabaticity parameter approaches zero for all $x$. This is more clearly seen in fig. \ref{fig2} which shows the location of the maximum ($x>0$) as function of $\Delta$ for three different couplings A.  

Note that the maxima are located at $x=\pm a$ for large detunings, which might be expected since that is where $|\partial g(x)|$ is maximal. For small detunings this is not the case, and a numerical check shows that for $\Delta\lesssim \sqrt{n}A/10$ the maxima deviate from $x=\pm a$.  Thus, for small $\Delta$ and within these approximations, this seems to confirm that a large coupling or a high excitation $n$ gives an adiabatic behaviour, which is contrary to the opposite case of $\Delta$ large. 

\subsection{Standing wave mode shape}\label{ssec52}

If the cavity field contains many spatial periods, the atom will see
the potential modulated by several periods $\lambda =\left( \frac{2\pi 
}{q}\right)$. The mode coupling then becomes%
\begin{equation}
g(x)=A\sin qx,  \label{a31}
\end{equation}%
which gives the condition for adiabaticity 
\begin{equation}
 \mathcal{A}^0(x)=\left\vert\left(\frac{p_0}{m}\right) \frac{q\Delta \sqrt{n}A\cos qx}{%
\left(\Delta ^{2}+4nA^{2}\sin ^{2}qx\right)^{3/2}}\right\vert \ll1.  \label{a32}
\end{equation}%
In the ordinary adiabatic limit of large detuning, this becomes %
\begin{equation}
\left\vert\left(\frac{p_0}{m}\right)\frac{q\sqrt{n}A}{ \Delta^2
 }\right\vert\ll1.  \label{a33}
\end{equation}%
We find, as we expect, the analogue of (\ref{a29}); the result improves with
increased detuning and lower cavity excitation.

The opposite limit $\Delta \rightarrow 0$ is more interesting and more subtle to analyze. Cavity QED systems operate in two different regimes; the optical \cite{optical} and microwave regime \cite{micro}. In the former case, the atomic wave packet may extend over several periods of the mode shape, and therefore sees a smeared periodic potential, while in the latter case $\lambda$ is assumed to be larger than the width $\Delta_x$ of the atomic wave packet. This results in very different adiabatic evolution in the small detuning limit, which will be seen in the next section. 

The adiabaticity parameter $\mathcal{A}^0(x)$ is a periodic function of $x$ and we may study it around one of the maxima, say $x=0$. For $x\approx0$ we find
\begin{equation}\label{adiabatapprox}
\mathcal{A}(x)\approx\frac{p_0}{2m\Delta}\frac{|C|}{\left(1+C^2x^2\right)^{3/2}},
\end{equation}
where $C=2q\sqrt{n}A/\Delta$. Note that this also gives the approximated adiabaticity parameter $\mathcal{A}^0(x)$ for a linear coupling $g(x)=Cx$. This is a sort of Lorenzian to the power of 3/2, which has a width that goes to zero when $\Delta\rightarrow0$ and an amplitude going to infinity in this limit. Thus, as $\Delta$ goes to zero, the function $\mathcal{A}^0(x)$ approaches zero for all $x$ except for $x=n\pi/q$, $n=...,-1,0,1,..$, where it becomes singular. Consequently we have 
\begin{equation}\label{limus}
\lim_{x\rightarrow n\pi}\lim_{\Delta\rightarrow0}\mathcal{A}^0(x)\neq \lim_{\Delta\rightarrow0}\,\lim_{x\rightarrow n\pi}\mathcal{A}^0(x).
\end{equation}
It is easy to show that the neglected higher order term $\partial^2\theta/\sqrt{\Delta^2+4G^2}$ is independent of the ordering of the limits in eq. (\ref{limus}). Therefor we conclude that eq.~(\ref{limus}) is still valid when the neglected term is included. In section \ref{sec7} we will show that the spread of the wave packet causes the adiabaticity parameter to be non-zero for other $x$ than $x=n\pi/q$ in the limit $\Delta\rightarrow0$.

\begin{figure}[ht]
\centerline{\includegraphics[width=8cm]{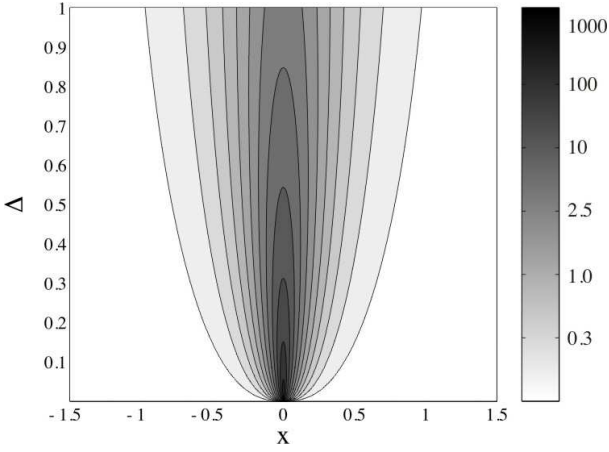}}
\caption[adiabaticpara2]{The adiabaticity parameter $\mathcal{A}^0(x)$ as function $x$ and $\Delta$ (scaled units) for the short optical wave length situation (\ref{a31}). It is seen that $\mathcal{A}^0(x)$ becomes singular at $x=0$ when $\Delta\rightarrow0$. The other parameters are $A=1$, $q=1$ and $p_0=2$. \label{fig3} }
\end{figure}

In fig. \ref{fig3} we show a plot of one of the maxima ($x=0$) for $\mathcal{A}^0(x)$, eq. (\ref{a32}), as function of $x$ and $\Delta$. As expected, the width increase with the detuning while the amplitude decrease. It is not clear if sharp or broad peaks of the parameter $\mathcal{A}^0(x)$ favour an adiabatic evolution, which will be examined in the next section. However, we may note that the integral of (\ref{adiabatapprox})
\begin{equation}
\int_{-\infty}^{+\infty}dx\,\mathcal{A}^0(x)=\left\vert\frac{p_0}{m\Delta}\right\vert
\end{equation} 
is independent of the parameter $C$, but depends of the detuning $\Delta$.

\section{Numerical investigations}\label{sec6}

The numerical studies are carried out by simulating the evolution of wave packets using the split operator method \cite{split}. The atomic wave packet will be taken to be in an initial {\it Gaussian bare state} in its upper level
\begin{equation}\label{initial}
\Phi(x,t=0)=\left[\begin{array}{c} \Phi_+(x)  \\ 0\end{array}\right]
\end{equation}
with the Gaussian
\begin{equation}
\Phi_+(x)=\frac{1}{\sqrt[4]{\pi\Delta_x^2}}\exp\left(-\frac{(x-x_0)^2}{2\Delta_x^2}\right)\exp(ip_0x).
\end{equation}  
Throughout the paper we use scaled dimensionless parameters in all numerical considerations, such that $\hbar=1$ and $m=1$. The 'exact' evolution is obtained by evolving (\ref{initial}) with the complete Hamiltonian (\ref{a5}), where we will use {\it Case} 1 with $\Lambda=\omega$, while the adiabatic state $\Psi^{ad}(x,t)=\left[\Psi_\uparrow^{ad}(x,t),\,\Psi_\downarrow^{ad}(x,t)\right]$, where the states $\Psi_{\uparrow,\downarrow}^{ad}(x,t)$ are defined in eq. (\ref{adstate}). The comparison between the two are measured by calculating the fidelity
\begin{equation}\label{fidelity}
F(t)=\int dx\,\Psi^{ad*}(x,t)\Phi(x,t).
\end{equation}
In the case of $p\approx p_0$ we choose to plot the fidelity as function of $\langle x\rangle\approx p_0t/m+x_0$, in order to get a better idea of where the fidelity changes along the mode $G(x)$. 

\subsection{Numerical treatment of a Gaussian mode}\label{ssec61}
\begin{figure}[ht]
\centerline{\includegraphics[width=8cm]{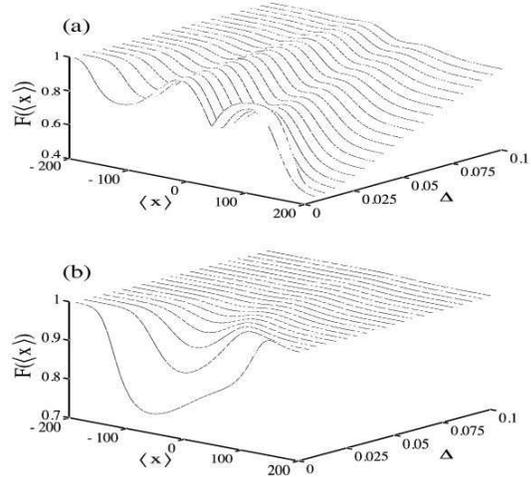}}
\caption[fidelity1]{This and the next figure \ref{fig5} show the fidelity (\ref{fidelity}) as function of $\langle x\rangle$ and $\Delta$ for the Gaussian mode shape (\ref{a27}) with $A=10$ (a) and $A=1$ (b). It is seen that the adiabaticity/fidelity is increased for smaller couplings $A$ when the detuning $\Delta$ becomes large. The other parameters are $\Delta_x=10$, $x_0=-200$, $p_0=5$ and $a=50$.    \label{fig4} }
\end{figure}

\begin{figure}[ht]
\centerline{\includegraphics[width=8cm]{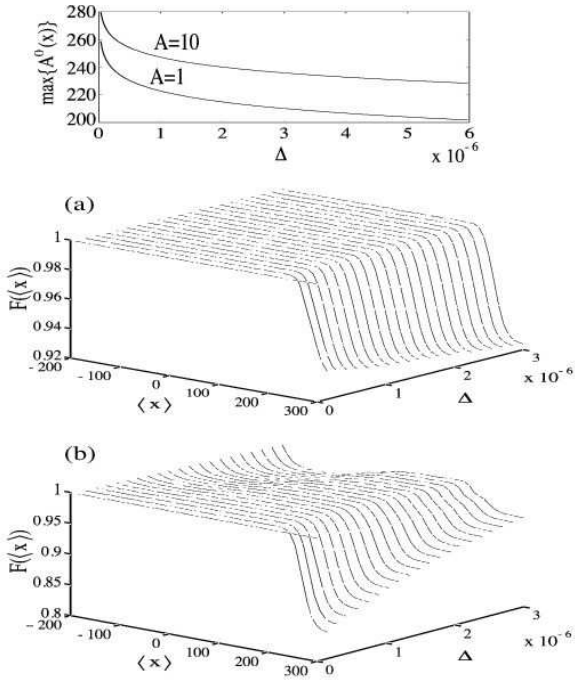}}
\caption[fidelity2]{The same as fig. \ref{fig4}, but for small detunings. The inset shows the location of the maximum of the adiabaticity parameter $\mathcal{A}(x)$ as function of the relevant detunings and the two coupling amplitudes $A=10$ (a) and $A=1$ (b). The wave packet is propagated to $x=300$ contrary to $x=200$ in the previous figure.  The reason is to see the drop in the fidelity around $x\approx200$ where $\mathcal{A}(x)$ has a maxima. The inset indicates that wave packets propagating with larger $\Delta$ will 'see' more of the maximum of $\mathcal{A}(x)$. The other parameters are as in fig. \ref{fig4}. \label{fig5} }
\end{figure}

In this section we investigate the adiabaticity criteria, derived in the previous section, for a Gaussian mode shape, while next section studies the periodic mode instead. We have dealt with the limiting cases of large and small detunings $\Delta$, and it was found that having an adiabatic evolution supports a small coupling, or photon excitation $n$, in the former, while for the latter it is the reverse case. In figs. \ref{fig4} and \ref{fig5}, we display the fidelity (\ref{fidelity}) for moderate and small detunings respectively. In the (a) plots the coupling strength is larger than in the (b) ones. The predicted result is clearly seen; the fidelity is improved for small $A$ in fig. \ref{fig4} and for large $A$ in fig. \ref{fig5}. In these plots we have chosen to plot the fidelity as a function of $\langle x\rangle$ rather than $t$, which has been done simply by transforming $\langle x\rangle\approx p_0t+x_0$. This should be justified since $p$ is approximately a constant of motion. The atomic wave packet starts at $x_0=-200$, outside the coupling pulse (width is here $a=50$), and then propagates across the pulse to $x\approx200$ in the first example and $x=300$ in the second. The reason why the packet is propagated longer in the second example is to show how the fidelity goes down when the wave packet enters the area, around $x=200$, where $\mathcal{A}^0(x)$ has its maximum. This is shown in the inset, which gives the location of the maximum ($x>0$) as function of $\Delta$ for the two coupling strengths. 

From figs. \ref{fig4} and \ref{fig5} we have seen that the expected result suggested by the lowest order of approximation of the adiabaticity criterion from the previous section. However, in the examples $p$ is almost perfectly a constant of motion, which justifies one of the approximations. The other one, coming from neglecting the higher order term $\partial^2\theta$,  will be considered in section \ref{sec7}.

\subsection{Numerical treatment of a standing wave mode}\label{ssec62}
As was pointed out in section \ref{sec5} for the standing wave mode, the limit $|\Delta|\ll G(x)$ brought about an adiabaticity parameter $\mathcal{A}^0(x)$ singular around the points $x=n\pi/q$ with $n$ an integer. At these points we can not have $|\Delta|\ll G(x)$ as $G(x)$ vanishes there. 

\begin{figure}[ht]
\centerline{\includegraphics[width=8cm]{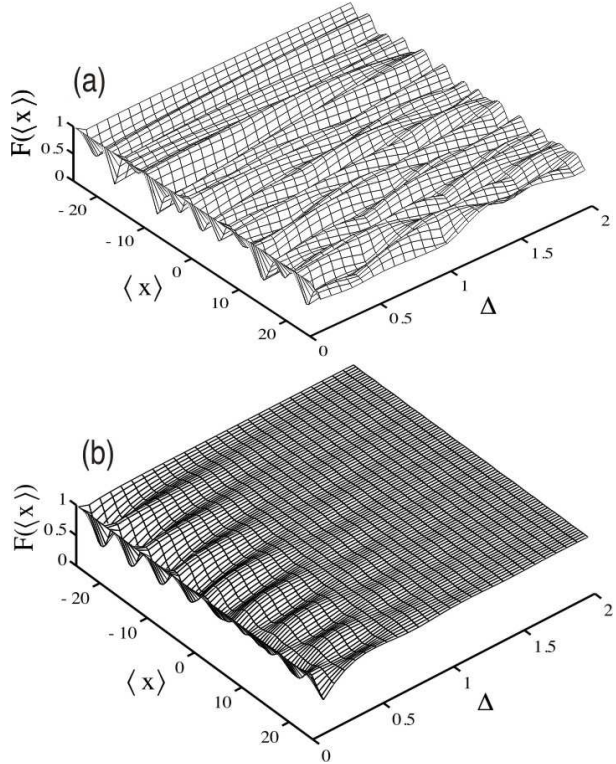}}
\caption[fidelity3]{The fidelity (\ref{fidelity}) as function of $\langle x\rangle$ and $\Delta$ for the short optical wave length mode shape (\ref{a31}) with $A=1$ (a) and $A=0.1$ (b). It is seen that the adiabaticity/fidelity is increased for smaller couplings $A$ when the detuning $\Delta$ becomes large. The other parameters are $\Delta_x=10$ (the width of the wave packet covers several wave lengths), $x_0=-25$, $p_0=5$ and $q=1$.    \label{fig6} }
\end{figure}

In the optical regime, the atomic wave packet is supposed to extend over several wave lengths, and consequently several $\delta$-like non-adiabatic singular couplings when $\Delta\rightarrow0$, and an adiabatic behaviour is not to be expected. The dynamics of wave packets evolving according to the Hamiltonian (\ref{a5}), with a coupling $g(x)=A\sin(qx)$ and a short optical wave length $\lambda\ll\Delta_x$ has been studied in \cite{jonas6}. The periodic Hamiltonian has a characteristic spectrum of energy bands $E_\nu(k)$ separated by forbidden gaps. Here $\nu$ is the band index and $k$ the quasi momentum. In large parameter ranges, including cases when $|\Delta|\ll A$, the atomic wave packet evolves as a 'free' particle with an effective group velocity and mass, both determined from the spectrum. Thus, it is tempting to expect that the 'freely' evolving Gaussian wave packet follows some kind of adiabatic behaviour. However, this turns out to be a false statement. The 'freely' evolving characteristic of the wave packet came from the fact that the initial Gaussian bare state (\ref{initial}) is approximately an eigenstate of the Hamiltonian in these parameter ranges, and since the spread of the wave packet is much larger than the wave length, the eigenenergy $E_\nu(k)$ may be expanded to second order around the initial quasi momentum $k_0$. It is a known fact that an initial  Gaussian wave packet evolving due to an energy $E=c_0+c_1p+c_2p^2$, for some constants $c_0$, $c_1$ and $c_2$, remains Gaussian. As it turns out, the initial Gaussian bare state (\ref{initial}) does not approximate an eigenstate of the adiabatic Hamiltonian (\ref{adH}) in the small detuning limit and will therefore not evolve as a 'free' particle under that Hamiltonain. The fidelity will therefor be small. However, in the large detuning limit $|\Delta|\gg G(x)$, adiabaticity is seen, as $\mathcal{A}^0(x)\ll1$ for all $x$. The above is shown in fig. \ref{fig6}, which displays the fidelity (\ref{fidelity}) as function of position and detuning. For large detunings the adiabaticity is seen and the fidelity approaches unity for increasing $\Delta$ and decreasing coupling amplitudes $A$.    

\begin{figure}[ht]
\centerline{\includegraphics[width=8cm]{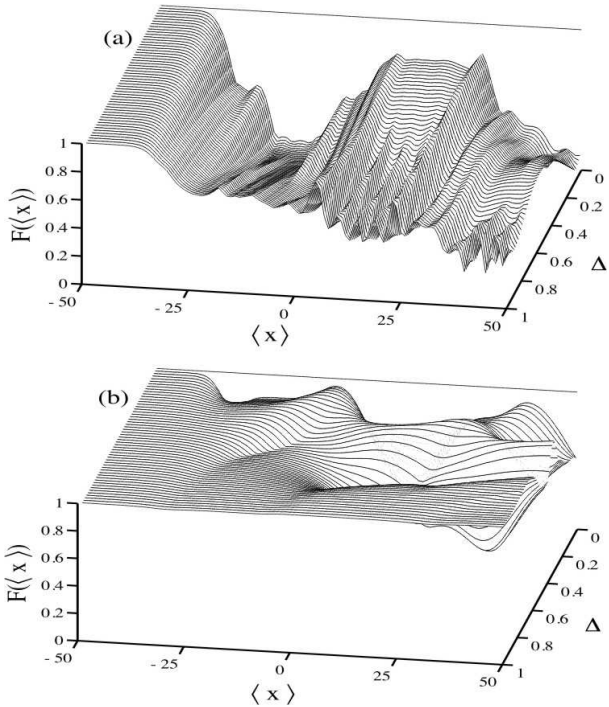}}
\caption[fidelity3]{The same as fig. \ref{fig6} but for a long wave length; $\lambda=20\pi$ and $\Delta_x=4$. In (a) $A=1$ and (b) $A=0.1$. Note that around $x=-10\pi,\,0,\,10\pi$, the fidelity changes the most. The initial position and momentum are $x_0=-50$ and $p_0=5$. 
\label{fig7} }
\end{figure}

When $\lambda>\Delta_x$ and $\Delta\rightarrow0$, the atom traverses intervals where $\mathcal{A}^0(x)\ll1$ for the whole extension of the wave packet and the fidelity (\ref{fidelity}) should stay fairly constant. When the wave packet approaches the singular points $x=n\pi/q$, a change in the fidelity is expected. For a given detuning and coupling amplitude, the width $\Delta_x$ and momentum $p$ roughly determine the rate of this change in the fidelity. As pointed out in the previous subsection \ref{ssec61}, it is not necessary that a drop in the fidelity is seen when the atom traverses a singularity. In fig. \ref{fig7} we show the fidelity for two different coupling strengths $A=1$ (a) and $A=0.1$ (b). Here $\lambda=20\pi$, while $\Delta_x=4$. It is seen that the main changes of the fidelity occurs at the points $x=n\pi/q$.   

\section{Higher order corrections to $\mathcal{A}_t$}\label{sec7}

In this section we will calculate numerically the adiabaticity parameter (\ref{adpara}) and compare with the approximated result of eq. (\ref{adparaapp}). We will also discuss the semiclassical time dependent model, which is obtained by the replacement $p\rightarrow p_0$ and $x\rightarrow p_0t/m+x_0$.

In the approximated result (\ref{adparaapp}), operators where replaced by semiclassical parameters and no averaging over the wave packet was carried out. Thus, the atom is seen as a point-like particle moving with well defined velocity and position. In order to take into account the quantum uncertainties we need to average over the wave packet. Note that the averaging is included already in eq. (\ref{a24}). If averaging over the fractions in (\ref{adpara}), the relative order of $x$ and $p$ must be specified, {\it e.g.} $(\partial\theta)p(\Delta_+-\Delta_-)^{-1}\neq(\Delta_+-\Delta_-)^{-1}(\partial\theta)p$. The separate wave packets $\Psi_\uparrow^{ad}(x,t)$ and $\Psi_\downarrow^{ad}(x,t)$ evolves differently, therefore $\mathcal{A}_t$ will depend on the initial condition $\Psi(x,0)$. The effect of averaging depends on the the shapes of the wave packets and the coupling, if $G(x)$ various considerably over the wave packet widths $\Delta_x$, the averaging has the effect of smearing out the variations. This is the case for a fast oscillating standing wave mode in an optical cavity, and therefor this case will be left out from this section and we only consider the microwave situation for the standing wave mode. We also note that if the opposite holds; $G(x)$ is almost constant over $\Delta_x$, then we have $\langle G(x)\rangle\approx G(\langle x\rangle)$, which holds for any smooth function $G$. 

\subsection{The parameter $\mathcal{A}_t$ for a Gaussian mode shape}\label{ssec71}
\begin{figure}[ht]
\centerline{\includegraphics[width=8cm]{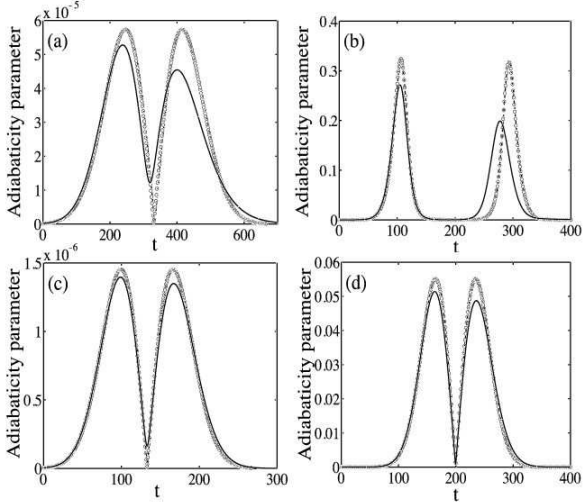}}
\caption[adparanum1]{These figures show the adiabaticity parameter (\ref{adpara}) calculated from wave packet simulations (solid line), the corresponding approximated result (\ref{adparaapp}) (dashed line) and the same with the $\partial^2\theta$ term included in (\ref{adparaapp}) (circles). In (a) and (b) the coupling amplitude is $A=100$ which makes $p$ vary considerably, while in (c) and (d) $A=1$ giving that $p\approx p_0$. The left plots show the large detuning situation $\Delta=10$, while $\Delta=0.05$ for the right ones. Note that the dashed and dotted lines are very similar, indicating that the main contribution comes from the first derivative of $\theta$. The remaining parameters are, (a): $p_0=0.6$, $x_0=-200$, (b): $p_0=1.5$, $x_0=-300$, (c): $p_0=1.5$, $x_0=-200$, (d): $p_0=1.5$, $x_0=-300$, and $a=50$ and $\Delta_x=10$.   \label{fig8} }
\end{figure}

The width of the wave packet is assumed to be narrower than the width of the Gaussian mode coupling. This means that the peaks seen in $\mathcal{A}^0(x)$ of fig. \ref{fig1} should not become considerably wider due to the averaging. The small broadening is expected to be more pronounced at later times, that is for the second maximum of the adiabaticity parameter, since the initially Gaussian wave packets $\Psi_{\uparrow,\downarrow}(x,t)$ spread with time. In fig. \ref{fig8} we give the results from numerical calculations of $\mathcal{A}_t$ (solid line), as defined in eq. (\ref{adpara}), together with the approximated adiabaticity parameter $\mathcal{A}^0(x)$ (dashed line), where $x=p_0t/m+x_0$. The circles displays $\mathcal{A}^0(x)$ with the neglected term $\partial^2\theta$ included. For the upper two plots, the momentum is small (compared to $A=100$) and therefore varies during the evolution, while in the lower ones we have $p\approx p_0$ ($A=1$), which is also seen by the better agreement between dotted/circles and solid lines. The left figures have a large detuning $\Delta=10$ while the others have $\Delta=0.05$. We note that the main contribution to the results comes from the term $\partial\theta p$ and not $\partial^2\theta$, as expected. The broadening of the peaks can also be seen.  

\subsection{The parameter $\mathcal{A}_t$ for a standing wave mode shape}\label{ssec72}

\begin{figure}[ht]
\centerline{\includegraphics[width=8cm]{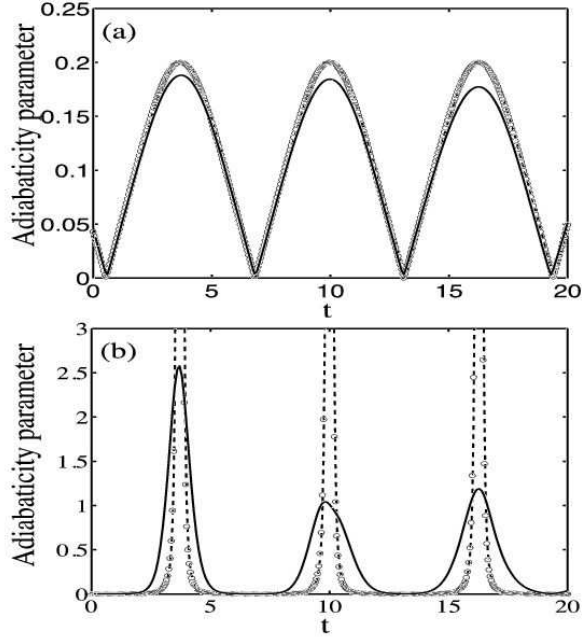}}
\caption[adparanum1]{ This shows the same as the previous fig. \ref{fig8}, but for the standing wave mode of eq. (\ref{a31}). Again, the dashed line displays the approximated result (\ref{adparaapp}) and the circles the same but with the $\partial^2\theta$ term added. The broadening of the peaks, coming from averaging over the atomic wave packet is clearly seen in (b) where the peaks are much more narrow. The parameters are, (a): $A=0.1$, $\Delta=0.5$, (b): $A=1$, $\Delta=0.05$, and for both plots $\Delta_x=4$, $x_0=-50$, $p_0=5$ and $q=0.1$. \label{fig9} }
\end{figure}

As already pointed out, when the wave length of the standing wave mode is shorter than the width of the wave packet the adiabaticity parameter (\ref{adpara}) will be smeared out due to the averaging over the wave packets. Thus, we only consider the case of a microwave field, where the extent of the wave packet may be smaller than the period of the mode. The approximated result (\ref{adparaapp}) indicated that for small detunings, $\mathcal{A}_t$ should be small for all $x$ except at the points $x=n\pi/q$ where it blowed up. However, the averaging will broaden these spikes. This is shown in fig. \ref{fig9}, which gives the results of $\mathcal{A}_t$ from wave packet propagation, compared with the analytically approximated results. The expected broadening is especially seen in the plot with narrow peaks, (a) $A=0.1$, $\Delta=0.5$, while in (b) $A=1$, $\Delta=0.05$. In (b) the momentum is not a good constant of motion; $\langle p\rangle_t\neq p_0$. We note that the main contribution comes from the term $(\partial\theta)p$ and not $\partial^2\theta$. 
    
\section{Relation to semiclassical time dependent Hamiltonians}\label{sec8}

In this section we investigate how one may derive effective time dependent models by using the adiabaticity parameter $\mathcal{A}_t$. 

Above we loosely argued, that for particles with a kinetic energy much larger than the interaction energy, we could derive an effective semiclassical time dependent model with $x=p_0t/m+x_0$. The Hamiltonian then reads
\begin{equation}\label{htime}
\tilde{H}=\left[\begin{array}{cc} \frac{\delta}{2} & G(t) \\ \\ G(t) & -\frac{\delta}{2}\end{array}\right],
\end{equation}
where $\sim$ indicates that it is an effective time dependent model. Here it is understood that $t$ and $x$ are related according to $x=p_0t/m+x_0$. 
Going to the adiabatic frame, by diagonalyzing $\tilde{H}$ with the corresponding unitary operator $\tilde{U}$ as in eq. (\ref{a11}), the Schr\"odinger equation becomes
\begin{equation}\label{timeadH}
i\frac{d}{dt}\left[\begin{array}{c} \tilde{\Psi}_\uparrow(t) \\ \tilde{\Psi}_\downarrow(t)\end{array}\right]=\left(\left[\begin{array}{cc} \tilde{\Delta}_+ & 0 \\ 0 & \tilde{\Delta}_-\end{array}\right]+i\frac{d\tilde{U}}{dt}\tilde{U}^{-1}\right)\left[\begin{array}{c} \tilde{\Psi}_\uparrow(t) \\ \tilde{\Psi}_\downarrow(t)\end{array}\right]. 
\end{equation}
Here the diagonal terms $\tilde{\Delta}_\pm$ are given by the corresponding expressions (\ref{a15}), and the adiabatic correction term is
\begin{equation}
\frac{d\tilde{U}}{dt}\tilde{U}^{-1}=\left[\begin{array}{cc} 0& \partial_t\tilde{\theta} \\ -\partial_t\tilde{\theta} & 0\end{array}\right].
\end{equation}
Demanding the off-diagonal term to be small compared with the 'distance' between the diagonal terms, gives the adiabaticity parameter
\begin{equation}\label{timeadpara}
\tilde{\mathcal{A}}_t=\left\vert\frac{\delta\sqrt{n}\partial g(x)}{\left(\delta^2+4ng^2(x)\right)^{3/2}}\left(\frac{p_0}{m}\right)\right\vert_{x=p_0t/m+x_0},
\end{equation}
which is identical to the approximate result (\ref{adparaapp}). This has several interesting consequences: \\ 

{\it i}) The analysis carried out in this paper concerns wave packet propagation, and since the approximate adiabaticity criterion we have defined is the same as the one obtained from time dependent models, much of the conclusions made in this paper are applicable to such time dependent models as well. \\ 

 {\it ii}) The constrain $\tilde{\mathcal{A}}_t\ll1$ is identical with the one standardly given in the literature
\begin{equation}\label{standard}
\left\vert\frac{\langle\tilde{\Psi}_\uparrow(t)|\partial_t|\tilde{\Psi}_\downarrow(t)\rangle}{\Delta_+(t)-\Delta_-(t)}\right\vert\ll1.
\end{equation}
Recently it has been argued that the criterion (\ref{standard}) does not, in general, guarantee adiabaticity \cite{adcond, adreplay}. The definition (\ref{adpara}) takes into account for the initial state being used, that is; the atom may start in any linear combination of upper and lower atomic states and it is apparent in the population weights $N_\uparrow^0$ and $N_\downarrow^0$. The corresponding definition for a time dependent model with instantaneous eigenvalues $E_n(t)$ and eigenstates $|n(t)\rangle$ would read
\begin{equation}
\tilde{\mathcal{A}}_t=\sum_n\sum_{m\neq n}|c_n^0|^2\left\vert\frac{\langle m(t)|\partial_t|n(t)\rangle}{E_m(t)-E_n(t)}\right\vert,
\end{equation}
where $|c_n^0|^2$ is the initial population of state $|n(0)\rangle$.
\\ 

{\it iii}) Deriving an effective time dependent model as above, is only justified when the two wave packets $\Psi_\uparrow^{ad}(x,t)$ and $\Psi_\downarrow^{ad}(x,t)$ follow nearly the same trajectory, since $x$ is replaced by a single quantity; $x=p_0t/m+x_0$. It must also be that the mode changes smoothly across the wave packet in order to neglect $\partial^2\theta$ and allow $\langle h(x,p)\rangle\approx h(\langle x\rangle,\langle p\rangle)$. Both these approximations neglect quantum mechanical corrections; the term $\partial^2\theta$ does not show up in the semiclassical models since the Schr\"odinger equation is first order in time, but second order in $x$. The quantum uncertainties in $x$ and $p$ are clearly ignored in $\mathcal{A}^0(x)$.  \\  
  
We now turn to the problem of deriving a time dependent Hamiltonian when $p$ is not a good constant of motion. This method is a kind of inverse problem. Given $\mathcal{A}_t$ we may construct a set of Hamiltonians $\tilde{H}(t)$ in the form of (\ref{htime}) that all give this particular adiabaticity parameter. Thus we solve for $G(t)$ from the equation
\begin{equation}\label{tmethod}
\frac{\delta\partial_tG(t)}{\left(\delta^2+4G^2(t)\right)^{3/2}}=\mathcal{A}_t,
\end{equation}
with solution
\begin{equation}\label{timecoup}
G(t)=\frac{f(t)}{\left(\delta^2+4f(t)\right)^{1/2}},
\end{equation}
where $f(t)=\int_{-\infty}^tdt'\,\mathcal{A}_{t'}$. This defines a set of time dependent Hamiltonians for a given adiabaticity parameter, where the coupling $G(t)$ and detuning $\delta$ are related as in (\ref{timecoup}). Note that this set of effective Hamiltonians is unique for one particular initial condition governing the $\mathcal{A}_t$. Thus, it gives an effective time dependent model for a certain Hamiltonian $H(x,p)$ and initial condition $\Psi(x,0)$. For a smoothly varying function $h(x)$, over the spread of the wave packets, we have $\langle h(x)\rangle\approx h(\langle x\rangle)$. Within this approximation, and neglecting the $\partial^2\theta$ term, the adiabaticity parameter (\ref{adpara}) becomes
\begin{equation}
\mathcal{A}_t=\frac{1}{2m}\left(N_\downarrow^0\left\vert\frac{ \partial\theta(x_\downarrow)p_\downarrow}{\Delta_+(x_\downarrow)}\right\vert+N_\uparrow^0\left\vert\frac{\partial\theta(x_\uparrow)p_\uparrow}{ \Delta_+(x_\uparrow)}\right\vert\right),
\end{equation}
where $x_i$, $p_i$, $i=\downarrow,\,\uparrow$ are the expectation values with the adiabatic wave functions $\Psi_{\downarrow,\uparrow}^{ad}(x,t)$. When these approximations are supposed to hold, it is reasonable to expect that we can replace the expectation values by their classical counterparts
\begin{equation}
\left\{\begin{array}{l} \partial_tp_\downarrow=\displaystyle{\frac{\partial\Delta_+(x_\downarrow)}{\partial x_\downarrow}} \\ \\ \partial_tx_\downarrow=\displaystyle{\frac{p_\downarrow}{m}}\end{array}\right.,\hspace{1cm}\left\{\begin{array}{l} \partial_tp_\uparrow=-\displaystyle{\frac{\partial\Delta_+(x_\uparrow)}{\partial x_\uparrow}} \\ \\ \partial_tx_\uparrow=\displaystyle{\frac{p_\uparrow}{m}}\end{array}\right..
\end{equation}
This is a way to derive a time dependent semiclassical model when $p$ is not a constant of motion and the wave packets traverse different trajectories. The error of such an effective Hamiltonian compared to exact ones is left for future research. It may be noted that one $\mathcal{A}_t$ defines a set of time dependent Hamiltonians of the form (\ref{htime}), and it is an open question how they are related apart from having the same adiabaticity constrains. For example, is there a particular detuning $\delta$ of the time dependent model such that the approximated effective Hamiltonian becomes an optimal approximated model? Note that if the approximation of replacing $p$ and $x$ by $p_0$ and $x=p_0t/m+x_0$ has been carried out, we note from (\ref{tmethod}), that we regain the result $G(t)=G(x)|_{x=p_0t/m+x_0}$, provided that we identify $\delta=\Delta$. As it is defined in the equation, (\ref{htime}) is not the most general two-level time dependent Hamiltonian, since the diagonal terms are assumed constant. If they are allowed to change with time, the method above becomes much more complicated. In general, to specify both $G(t)$ and $\delta$ we need, in addition to eq. (\ref{tmethod}), a second equation.  

\section{Conclusion}\label{conclu}

In this paper we have discussed the concept of adiabaticity within cavity QED models. To do so, the original Jaynes-Cummings model, describing the atom-cavity field interaction, has been modified by taking into account, both the center-of-mass position and momentum of the atom. Two different cases have been studied: The atom sees a Gaussian or a standing wave mode shape. Numerical simulations have been used to measure how well the atomic wave packet follows the cavity mode shape adiabatically. These results have been compared with approximated analytically derived criteria for the adiabaticity. Good agreement between the two has been found. 

The large detuning case, generally called the adiabatic limit, implies adiabatic following for any mode shape. More interesting is that adiabaticity is, for the Gaussian mode, obtained in the opposite limit as well, the small detuning case. Contrary to the large detuning situation, here a large atom-field coupling or a high cavity excitation improves adiabaticity. This is discussed in detail and it turns out that for the standing wave mode case, this kind of adiabatic limit does not exist.

The approximations in the derivation of the analytic adiabaticity criterion have been checked by comparing it with the more correctly defined measure, which is an average over the atomic wave packet as it traverse the cavity field. The averaging tends to smear out rapid variations of the criterion.

Finally, we have discussed how the wave packet propagation problem may be transformed into an effective time dependent semiclassical two-level problem. The most simple situation is when $p$ is roughly a constant of motion for both of the internal adiabatic wave packets, and one replaces $p\rightarrow mv$ and $x\rightarrow vt+x_0$ in the original Hamiltonian. The procedure is not straightforward when $p$ is not a constant of motion and the two wave packets traverse different trajectories. A systematic way to derive a set of semiclassical Hamiltonians is presented in these cases. It is suggested when such a method is supposed to work, but no deeper analysis of the validity is carried out, and is left for future investigations. 

The paper has been devoted to cavity QED systems, but it should be mentioned that the analysis can be generalized to a selection of other systems, such as: trapped ions, molecular coherent processing, quantum electronics, BEC, and also time dependent models. It is also straightforward to consider multi-level atoms and/or multi-modes Jaynes-Cummings models.

\pagebreak


\begin{thebibliography}{99}

\bibitem{jc} E. T. Jaynes, and F. W. Cummings, Proc. IEEE {\bf 51}, 89 (1963). B. W. Shore, and P. L. Knight, J. Mod. Opt. {\bf 40}, 1195 (1993).

\bibitem{adiabat} M. Born, and V. Fock, Z. Phys. {\bf 51}, 165 (1928). A. Messiah, {\it Quantum Mechanics, vols. 1 and 2}, Wiley (1962).

\bibitem{berry} M. V. Berry, Proc. Roy. Soc. Lond. A {\bf 392}, 45 (1984). A. D. Shapere, and F. Wilczek, {\it Geometric phases in physics}. (World Scientific, 1989).

\bibitem{adcomp} E. Farhi, J. Goldstone, and M. Sipser, quant-ph/0001106. 

\bibitem{adcond} K. P. Marzlin, and B. C. Sanders, phys. Rev. Lett. {\bf 93}, 160408 (2004). D. M. Tong, K. Singh, L. C. Kwek, and C. H. Oh, Phys. Rev. Lett. {\bf 95}, 110407 (2005).

\bibitem{adreplay} M. S. Sarandy, L. -A. Wu, and D. A. Lidar , Quant. Info. Proc. {\bf 3}, 331 (2004). Z. Wu, and H. Yang, Phys. Rev. A {\bf 72}, 012114 (2005). A. K. Pati, and A. K. Rajagopal, quant-ph/0405129. R. MacKenzie, E. Marcotte, and H. Paquette, quant-ph/0510024. S. Duki, H. Mathur, and O. Narayan, quant-ph/0510131. T.~V\'ertsi, and R.~Englman, quant-ph/0511141. 

\bibitem{adel1} L. A. Lugiato, P. Mandel, and L. M. Narducci, Phys. Rev. A {\bf 29}, 1438 (1984). C. C. Gerry, and J. H. Eberly, Phys. Rev. A {\bf 42}, 6805 (1990).

\bibitem{adel2} A. B. Klimov, and L.L. Sanchez-Soto, Phys. Rev. A {\bf 61}, 063802 (2000). A. B. Klimov, L. L. Sanchez-Soto, A. Navarro, and E. C. Yustas, J. Mod. Opt. {\bf 49}, 2211 (2002).

\bibitem{born} M. Born, and R. Oppenheimer, Ann. der Phys. {\bf 84}, 30 (1927). H. Lefebvre-Brion, and R. W. Field, {\it Pertubations in the spectra of diatomic molecules}, (Academic Press, 1986).

\bibitem{jonas1} J. Larson, and S. Stenholm, J. Mod. Opt. {\bf 50}, 1663 (2003).

\bibitem{jonas2} R. R. Schlicher, Opt. Comm. {\bf 70}, 97 (1988). J. Larson, and S. Stenholm, J. Mod. Opt. {\bf 50}, 2705 (2003).
 
\bibitem{slow}B. G. Englert, J. Schwinger, A. O. Barut, and M. O. Scully, Euorophys Lett. {\bf 14}, 25 (1991). G. M. Meyer, M. O. Scully, and H. Walther, Phys. Rev. A \textbf{56}, 4142 (1997). R. Arun, and G. S. Agarwal, Phys. Rev. A {\bf 64}, 065802 (2001). T. Bastin, and J. Martin, Phys. Rev. A {\bf 67}, 053804 (2003).

\bibitem{jonas6} J. Larson, J. Salo, and S. Stenholm, Phys. Rev. A {\bf 72}, 013814 (2005).


\bibitem{optical} C. J. Hood, T. W. Lynn, A. C. Doherty, A. S. Parkins, and H. J. Kimble, Science {\bf 287}, 1447 (2000). P. W. H. Pinkse, T. Fisher, P. Maunz, and G. Rempe, Nature {\ 404}, 365 (2000). 

\bibitem{micro} S. Haroche in {\it Quantum entanglement and information processing}, edited by D. Asteve, J.- M. Raimond, and J. Dalibard, (Elsevier, 2004). D. Meschede, H. Walther, and G. MŸller, Phys. Rev. Lett. {\bf 54}, 551 (1985). 

\bibitem{split} M. D. Fleit, J. A. Fleck, and A. Steiger, J. Comp. Phys. {\bf 47}, 412 (1982).

 \end{thebibliography}
\end{document}